\documentstyle[12pt]{article}
\newlength{\extraspace}
\newlength{\extraspaces}
\newcommand{\be}{\begin{equation}
\addtolength{\abovedisplayskip}{\extraspaces}
\addtolength{\belowdisplayskip}{\extraspaces}
\addtolength{\abovedisplayshortskip}{\extraspace}
\addtolength{\belowdisplayshortskip}{\extraspace}}
\newcommand{\ee}{\end{equation}}

\newcommand{\ba}{\begin{eqnarray}
\addtolength{\abovedisplayskip}{\extraspaces}
\addtolength{\belowdisplayskip}{\extraspaces}
\addtolength{\abovedisplayshortskip}{\extraspace}
\addtolength{\belowdisplayshortskip}{\extraspace}}
\newcommand{\ea}{\end{eqnarray}}

\newcommand{\figuur}[2]{
\begin{figure}
\vspace{#1mm}
\begin{center}
\setlength{\unitlength}{1.1mm}
\raisebox{-20\unitlength}
{\mbox{\begin{picture}(70,65)(-35,-35)
\thicklines
\put(-30,0){\line(1,1){30}}
\put(0,30){\line(1,-1){30}}
\put(30,0){\line(-1,-1){30}}
\put(0,-30){\line(-1,1){30}}
\thinlines
\put(-12,12){\vector(-1,1){6}}
\put(12,12){\vector(1,1){6}}
\put(-18,-18){\vector(1,1){6}}
\put(18,-18){\vector(-1,1){6}}
\put(-8,16){\vector(-1,1){6}}
\put(8,16){\vector(1,1){6}}
\put(-14,-22){\vector(1,1){6}}
\put(14,-22){\vector(-1,1){6}}
\put(-16,8){\vector(-1,1){6}}
\put(16,8){\vector(1,1){6}}
\put(-22,-14){\vector(1,1){6}}
\put(22,-14){\vector(-1,1){6}}
\put(-10,-10){\makebox(0,0){$\Pip$}}
\put(10,-10){\makebox(0,0){$\Pim$}}
\put(-10,9){\makebox(0,0){$\Pom$}}
\put(10,9){\makebox(0,0){$\Pop$}}
\put(-9,-30){\makebox(0,0){${\cal I}^-$}}
\put(-9,30){\makebox(0,0){${\cal I}^+$}}
\put(-8,28){\line(1,0){10}}
\put(-8,-28){\line(1,0){10}}
\put(-8.5,-28.5){\line(1,1){3.5}}
\put(-8.5,28.5){\line(1,-1){3.5}}
\end{picture}}}
\parbox{13cm}{\small #2}
\end{center}
\end{figure}}

\newcommand{\asympt}[1]{
\begin{figure}
\begin{center}
\setlength{\unitlength}{.75mm}
\raisebox{-40\unitlength}
{\mbox{\begin{picture}(80,66)(-40,-40)
\thicklines
\put(-30,0){\line(1,1){30}}
\put(0,30){\line(1,-1){30}}
\put(30,0){\line(-1,-1){30}}
\put(0,-30){\line(-1,1){30}}
\thinlines
\put(-18,-17){\makebox(0,0){${\cal I^-}$}}
\put(-18,17){\makebox(0,0){${\cal I^+}$}}
\put(19,-17){\makebox(0,0){${\cal I^-}$}}
\put(20,17){\makebox(0,0){${\cal I^+}$}}
\put(0,0){\makebox(0,0){$M$}}
\put(-37,5){\makebox(0,0){$\partial M$}}
\put(-33,5){\vector(1,0){8}}
\end{picture}}}
\parbox{13cm}{\small #1}
\end{center}
\end{figure}}

\newcommand{\Xm}{X^{{}^{\! -}}}

\newcommand{\xm}{x^{{}^{\! -}}}

\newcommand{\oneshock}[1]{
\begin{figure}
\begin{center}
\setlength{\unitlength}{0.90mm}
\raisebox{-40\unitlength}
{\mbox{\begin{picture}(80,60)(-35,-35)
\thicklines
\put(-22,22){\line(1,-1){45}}
\put(10,-10){\vector(-1,1){1}}
\thinlines
\put(3,-3){\line(-1,-1){20}}
\put(-3,3){\line(1,1){20}}
\put(-6,-12){\vector(1,1){1}}
\put(6,12){\vector(1,1){1}}
\put(-12,-9){\makebox(0,0){{\mbox{\footnotesize ${\xm\!\! = \! \Xm}$}}}}
\put(20,7){\makebox(0,0){{\mbox{\footnotesize ${\xm\!\! = \! \Xm\!\! +\!
p_1^-\!
 f(y,y_1)}$}}}}
\put(16,-9){\makebox(0,0){${p_1^-}$}}
\put(0,18){\makebox(0,0){{${\mbox {\small geodesic}}$}}}
\put(8,18){\vector(1,0){4}}
\end{picture}}}
\parbox{13cm}{\small #1}
\end{center}
\end{figure}}

\newcommand{\newsection}[1]{
\vspace{15mm}
\pagebreak[3]
\addtocounter{section}{1}
\setcounter{equation}{0}
\setcounter{subsection}{0}
\setcounter{footnote}{0}
\begin{center}
{\large \thesection. #1}
\end{center}
\nopagebreak
\medskip
\nopagebreak}

\newcommand{\newsubsection}[1]{
\vspace{1cm}
\pagebreak[3]

\addtocounter{subsection}{1}
\noindent{ \sc \thesubsection. #1}
\nopagebreak
\vspace{2mm}
\nopagebreak}

\newcommand{\ie}{{\it i.e}.\ }
\newcommand{\eg}{{\it e.g}.\ }
\newcommand{\hf}{{\textstyle{1\over 2}}}
\newcommand{\tr}{{\rm tr}}
\newcommand{\is}{\! & \! = \! & \!}
\newcommand{\nonu}{\nonumber \\[1.5mm]}
\newcommand{\ab}{{\alpha\beta}}

\newcommand{\cd}{{\gamma\delta}}
\newcommand{\ppar}{{{}_{\! /\! /}}}
\newcommand{\pper}{{{}_{\! \perp}}}
\newcommand{\Xbar}{{\overline{X}}}
\newcommand{\Pip}{ P^{in}_{+}}
\newcommand{\Pim}{ P^{in}_{-}}
\newcommand{\Pop}{ P^{out}_{+}}
\newcommand{\Pom}{ P^{out}_{-}}
\newcommand{\rth}{\mbox{\raisebox{-.1ex}{$\sqrt{h}$}}}
\newcommand{\rtg}{\mbox{\raisebox{.2ex}{$\sqrt{g}$}}}

\begin{document}
\addtolength{\baselineskip}{.7mm}

\thispagestyle{empty}
\begin{flushright}
{\sc PUPT}-1279\\
{\sc IASSNS-HEP}-91/60\\
September 1991
\end{flushright}
\vspace{.3cm}

\begin{center}
{\large\sc{Scattering at Planckian Energies}}\\[13mm]

{\sc  Herman Verlinde}\footnote{Supported by NSF Grant PHY90-21984.}\\[3mm]
{\it Joseph Henry Laboratories\\[2mm]
Princeton University\\[2mm]
Princeton, NJ 08544} \\[.4cm]
{ and}\\[.4cm]
{\sc Erik Verlinde}\footnote{Supported by the W.M. Keck Foundation.}
\\[3mm]
{\it School of Natural Sciences\\[2mm]
Institute for Advanced Study\\[2mm]
Princeton, NJ 08540}
\\[8mm]

{\sc Abstract}
\end{center}

\noindent
We give a systematic analysis of forward scattering in 3$+$1-dimensional
quantum
gravity, at center of mass energies comparable or larger than the Planck
energy. We show that quantum gravitational effects in this kinematical regime
are described by means of a topological field theory. We find that the
scattering amplitudes display a universal behaviour similar to
two dimensional string amplitudes, thereby recovering results obtained
previously by 't Hooft. Finally, we discuss the two-particle process in some
detail.

\vfill

\newpage
\newsection{Introduction}

\noindent
Quantum gravitational effects grow large when the length and time scales of
the physical process of interest fall below the Planck length $\ell_{pl}$.
This well-known fact has the following, necessarily heuristic,
explanation.  If we decide to keep our coordinates dimensionless, the metric
$g_{\mu\nu}$ has dimension (length)$^2$. This dimensionality can be made
explicit by introducing some scale $L$, which we may take to be some length
characteristic of the physical situation we wish to consider, and write
\be
\label{metscale}
g_{\mu\nu} = L^2 \hat{g}_{\mu\nu},
\ee
where $\hat{g}_{\mu\nu}$ is dimensionless. The Einstein action
\be
\label{einstein}
S_E[g] = \int \! \rtg R
\ee
has the dimension of (length)$^2$, that is
\be
\label{actscale}
S_E[L^2\hat{g}] = L^2 S_E[\hat{g}] .
\ee
Now, although in the quantum theory we are supposed to integrate over all
metrics, it seems physically reasonable to suppose that in this integral only
those metrics are relevant whose size is comparable to that of the physical
system. In other words, we may reasonably assume that the rescaled
metric $\hat{g}_{\mu\nu}$ is typically of order 1. With this assumption, we
can now indeed conclude from (\ref{actscale}) that, since $S_E$ is normalized
by Newton's constant $G = \ell_{pl}^2$, the dimensionless coupling constant
determining the strength of the quantum gravitational effects is given by the
 ratio
\be
g(L) = {\ell_{pl} \over L}
\ee
of the Planck length $\ell_{pl}$ and the characteristic length $L$.

Because of this reasoning, it is generally believed that to predict
scattering processes in which the center of mass energy exceeds
the Planck energy, the full theory of quantum gravity (or perhaps
string
theory) has to be invoked. However, this is not necessarily true, since a
scattering
process between two particles is characterized by two energies,
\ie the center of mass energy and the momentum transfer, and
even if the former is very large, the latter can be small. Therefore, in
the physical context of very high energy collisions, we can regard quantum
gravity as a theory with {\it two} coupling constants.

This paper is an attempt to exploit this fact to start a systematic
study of high energy scattering in quantum gravity, in the limit where
the momentum transfer is small compared to the Planck energy.
This study was motivated by recent work by 't Hooft, who argued that
in this kinematical regime quantum gravity can be treated semi-classically,
and furthermore displays a remarkable formal resemblance to string
theory \cite{thooft1,thooft2}. Both results will be confirmed by our
analysis. We will explain why
at small momentum transfer indeed all physical modes of the gravitational
field become classical, except for two longitudinal modes. We then show
that the quantum
dynamics of these modes is described by means a topological field theory,
which can be characterized as a two-dimensional topological sigma model
with the group of diffeomorphisms in 1$+$1 dimensions as target space.
We will further study the interaction of this model with ultra-energetic
particles and show
that the amplitudes describing the forward scattering of these particles can
be rewritten in a form similar to 1+1 dimensional string amplitudes.
We will end with a discussion of the two-particle process.

\newsection{The Einstein Action with Two Coupling Constants}

To characterize the scattering processes that we are interested in,
let us imagine that some day the technology will be available to build a
linear accelerator that can accelerate elementary particles, say electrons,
to energies comparable or larger than the Planck energy. For concreteness,
let us choose our $x$ axis along this accelerator, and introduce the notation
$x^\alpha \equiv (t,x)$ for the longitudinal and
$y^i \equiv (y,z)$ for the transversal coordinates. Thus the momenta
of the particles are of Planckian magnitude in the $x^\alpha$-plane,
while the perpendicular momenta $p_i$ are negligible.
Now, since the transversal cross sections of the electron beams used in our
Gedanken experiments are necessarily very large compared to the Planck length,
in practice all collisions between the left- and right-moving electrons
have relatively very large impact parameters. Hence we are dealing
almost exclusively with forward scattering.

The physics of these forward scattering processes is characterized by two
length
scales of very different order of magnitude, namely $\ell_\ppar$, the typical
longitudinal wavelength of the particles, and $\ell_\pper$, the
characteristic length of the impact parameter. Hence, we can form the two
dimensionless ratios
\ba
g_\ppar \! \! & \! \equiv \! & \!\! {\ell_{pl}\over \ell_\ppar } \sim 1 \nonu
g_\pper \! \! & \! \equiv \! & \!\!  {\ell_{pl}\over \ell_\pper} <\!\! < 1 .
\ea
We will now show that, in this regime, quantum gravity indeed naturally splits
into two sectors,  one strongly coupled with $g_\ppar$,
and one weakly coupled with $g_\pper$.

\newsubsection{Splitting the Einstein Action}

Given the above geometrical setup, it is natural to make the following
gauge choice
\be
\label{twotwo}
g_{\mu\nu} = \left( \begin{array}{cc} g_\ab & 0 \\ 0 & h_{ij}
\end{array} \right).
\ee
Here $g_\ab$ has the Lorentzian signature, and $h_{ij}$ is Euclidean.
It is evident that locally the metric can always be brought in this form.
This gauge may look somewhat unfamiliar, and indeed, it is easy to see that
it is not suitable for doing perturbative calculations around an
(asymptotically) flat background.
Our interest, however, is in doing non-perturbative calculations in $g_\ppar$,
and for this
purpose it is sufficient to know that the above gauge choice is non-singular
around sufficiently generic metrics. Further, in general
there may be global obstructions to choosing the metric of the above form, but
these
do not occur in our situation, because we assume that the topology in the
$x^\alpha$ and $y^i$ direction is trivial.

The physical motivation for choosing the gauge (\ref{twotwo})
is that in the kinematical regime of interest, the characteristic scale
$\ell_\ppar$ of $g_\ab$ is of the order of $\ell_{pl}$, while the typical
length
$\ell_\pper$ of $h_{ij}$ is much larger.  Thus, following the line of argument
given in the Introduction, let us introduce two dimensionless metrics
$\hat{g}_\ab$ and $\hat{h}_{ij}$ via
\ba
\label{ghscale}
g_\ab \is \ell_\ppar^2 \, \hat{g}_\ab,
\nonu
h_{ij} \is \ell_\pper^2 \, \hat{h}_{ij}.
\ea
We again will make the plausible assumption that $\hat{g}$ and
$\hat{h}$ are of order 1.

\pagebreak

Now, a straightforward calculation, outlined in the Appendix,
shows that the Einstein action in the gauge (\ref{twotwo}) splits into two
terms
\be
S_E[g] =
S_\ppar[g,h] + S_\perp[h,g],
\ee
given by
\be
\label{s1}
S_\ppar[g,h] = \int \!\rtg \, \Bigl(\rth R_h +{\textstyle{1\over 4}}
 \sqrt{h} h^{ij}
\partial_i g_\ab
\partial_j g_\cd \epsilon^{\alpha\gamma}\epsilon^{\beta\delta}\Bigr), \\[1mm]
\ee
and
\be
\label{s2}
S_\perp[h,g] = \int \!  \rth \, \Bigl(\rtg R_g  +
{\textstyle{1\over 4}} \sqrt{g} g^\ab
\partial_\alpha h_{ij} \partial_\beta h_{kl}
 \epsilon^{ik}\epsilon^{jl}\Bigr).\\[2mm]
\ee
The crucial property separating both terms is that they behave
differently under the constant rescaling (\ref{ghscale}) of $g_\ab$ and
$h_{ij}$, namely\footnote{The
$\epsilon$-symbols in
(\ref{s1}) and (\ref{s2}) contain a factor of $(\sqrt{g})^{-1}$ resp.
$(\sqrt{h})^{-1}$, so that they transform as tensors.}
\ba
\label{rescale1}
S_\ppar[\, \ell_\ppar^2\,  \hat{g},\,  \ell_\pper^2\,  \hat{h}\, ]
\is  \ell_\ppar^2 \, S_\ppar[\, \hat{g},\, \hat{h}], \nonumber\\[3.5mm]
\label{rescale2}
S_\perp[\, \ell_\pper^2 \, \hat{h}, \, \ell_\ppar^2 \, \hat{g} \, ] \is
\ell_\pper^2\, S_\perp[\, \hat{h},\, \hat{g}].
\ea
Thus, after we make $S_E$ dimensionless by dividing by Newton's constant
$G=\ell_{pl}^2$, we see that the $S_\ppar$ part of the
Einstein action is strongly coupled, with coupling $g_\ppar =
\ell_{pl}/\ell_\ppar$, whereas the coupling constant $g_\pper =
\ell_{pl}/\ell_\pper$ of the other term $S_\perp$ is very small.
We arrive therefore at the important conclusion that
{\it as far as $S_\perp$ is concerned we are in the classical regime}.
In the following we will exploit this simplification of the dynamics of
quantum gravity to analyze the theory to leading order in $g_\pper$, while
keeping track of the full non-perturbative dependence on $g_\ppar$.

To complete the picture we should include the ghost action
associated with our gauge choice $g_{i\alpha} \! = \! 0$.
Under an infinitesimal coordinate transformation we have
$\delta g^{i\alpha} = h^{ij} \partial_j\xi^\alpha +
g^{\ab}\partial_\beta \xi^i$, so we find that the ghost action
consists of the two terms
\ba
\label{sgh}
S_{gh}[b,c] \is S_\ppar[b,c] + S_\perp[b,c] \nonumber \\[3mm]
S_\ppar[b,c]  \is \int \! \rtg \, \sqrt{h}h^{ij}\,
b_{i\alpha} \partial_j c^\alpha, \\[2.5mm]
S_\perp[b,c] \is \int \! \rth \, \sqrt{g} g^\ab \, b_{i\alpha}
\partial_\beta c^i. \nonumber
\ea
\nopagebreak
Again we observe that after the rescaling (\ref{ghscale})
the first term $S_\ppar$ is strongly coupled and the second $S_\pper$ weakly.

\pagebreak

\newsubsection{Isolating the Relevant Gravitational Modes}

It is instructive to make an inventory of the different degrees of freedom
of the gravitational field in relation with the above decomposition of the
action. As usual in quantum gravity we have, after gauge-fixing, 6 remaining
components of the metric and 4 anti-commuting ghost variables. The ghosts
eliminate the 4 longitudinal modes of the metric via the BRST-invariance of the
action and effectively impose the `Gauss law' constraints
\be
R_{i\alpha}=0.
\ee
This leaves 2 physical degrees of freedom constituting the gravitational
radiation modes.
A useful parametrization of the metric variables, which manifestly exhibits
the different modes, is
\ba
\label{param}
g_\ab \is e^{\phi(X,Y)} \, \partial_\alpha X^a \partial_\beta X^b \,
 \eta_{ab}
\\[3mm]\nonumber
h_{ij} \is e^{\chi(X,Y)}\, \partial_i Y^p \partial_j Y^q  \, \delta_{pq}.
\ea
Here $X^a(x,y)$ and $Y^p(x,y)$ are the diffeomorphisms that
relate $g_\ab$ resp. $h_{ij}$ to the diagonal metrics. As such, they
are paired with the ghost fields $c^\alpha$ resp.\ $c^i$ via the BRST symmetry.
Hence they are the longitudinal modes and {\it locally} decouple from
physical processes; as we will see, however, their global variations are
important observables. The conformal factors $\phi$ and $\chi$ are physical,
BRST-invariant fields and represent the two transversal components
of the graviton.

To see which of these modes are physically relevant in our context,
let us consider the coupling to matter.
In general, we can include matter by adding a source term to the
Einstein action
\be
S_E[g] \rightarrow S_E[g] + S_m[g].
\ee
This source term represents the stress-energy of the matter particles,
of which the transversal components $T_{ij}$ are assumed to be negligible
compared to longitudinal components
\be
\label{tab}
T_\ab = {1\over \sqrt{g}}{\delta S_m[g]\over \delta g^\ab}.
\ee
Here $T_\ab$ is effectively traceless, provided the rest-masses of the matter
particles are very small compared to the Planck mass. The remaining two
 components
of $T_\ab$ can be identified with the momenta of the particles, which are
either
left- or right-moving. We see therefore that the matter stress energy only
 couples
to the off-diagonal variations of $g_\ab$, which are parametrized by the fields
 $X^a$.

An important property of $S_\pper$ is that it is invariant under
reparametrizations $x^\alpha \rightarrow \tilde{x}^{\tilde{\alpha}}(x,y)$.
As a consequence, one finds that upon inserting the parametrization
(\ref{param}) into (\ref{s1}) and (\ref{s2}) that $S_\pper$ does not
depend on $X^a$. Now, since $g_\pper\! \! <\!\! <\! 1$,
the functional integral over metrics is strongly peaked near the space
of field configurations that minimize $S_\perp$, and thus
to leading order in $g_\pper$, we can restrict the configuration space
of gravity to this vacuum subspace. Combined with the above observation
that the matter only couples to the $X^a$ fields, this shows that all
other variables can be treated classically. The $X^a$ are therefore
the relevant quantum mechanical modes, since their dynamics is dictated purely
by the strongly coupled $S_\ppar$ piece of the action.

\newsection{The Topological Effective Theory.}

The restriction to the vacuum subspace of $S_\pper$ eliminates
the local gravitational degrees of freedom:  only the global variations
of the remaining longitudinal modes $X^a$ can be physically relevant.
In this section we will construct the topological field theory that
describes their dynamics.

{}~From the above discussion we learn that, to leading order in $g_\pper$, we
 can
restrict ourselves to the special class of vacuum field configurations such
that
\be
S_\perp[h,g] =  0.
\ee
This leads to the equations
\ba
\partial_\alpha h_{ij} \is 0  \ \ \ \nonu
R_g \is 0, \ \ \
\ea
which single out the vacuum solutions in which no graviton waves propagate.
These solutions are of the general form
\ba
\label{vacuum}
\ \ \ h_{ij} \is h_{ij}(y) \nonu
\ \ \ g_\ab \is \eta_{ab} \partial_\alpha X^a \partial_\beta X^b
\ea
The variables $X^a$ are maps of the two-dimensional $x$-plane onto itself
and vary in the transversal $y$ directions. Here the transversal metric
$h_{ij}(y)$ is allowed to be arbitrary and will be treated as a fixed classical
background.

Along with the metric, the ghost fields are also constrained to solve the
equation of motion of the weakly coupled $S_\perp$ piece of the ghost action
(\ref{sgh}).
This equation for the anti-ghost, $\,\nabla^\alpha b_{i\alpha} = 0 $, is solved
 by
\be
\label{ghredef}
\ \ \ b_{i\alpha} = \epsilon_\ab \partial^\beta b_i,
\ee
where the new anti-ghosts $b_i$ are unrestricted field variables.
Substituting the parametrizations (\ref{vacuum}) and (\ref{ghredef})
into the Einstein action gives the effective action for the new
variables $X^a$, $c^\alpha$ and $b^i$.
\newcommand{\Vi}{{V_i}}

To write this reduced action, it is convenient to introduce the
vector fields $\Vi^\alpha$ via the relation
\be
\label{parX}
\partial_i X^a +\Vi^\alpha \partial_\alpha X^a=0.
\ee
These variables $\Vi^\alpha$ describe the flow of the $X^a$ fields in the
$y$-direction, and are analogous to the fluid velocity in fluid mechanics.
The transversal derivatives of the metric can be expressed in terms of
$\Vi^\alpha$ as
\be
\label{variation}
\partial_i g^\ab = \nabla^\alpha \Vi^\beta
+ \nabla^\beta \Vi^\alpha.
\ee
Here and in the following we keep the $X^a$ dependence implicit in $g_{\ab}$
and $\Vi^\alpha$.
After substituting this into (\ref{s1}) and using the identity
$\epsilon_{\alpha\delta} \epsilon_{\beta\gamma}=
\epsilon_{\alpha\gamma}\epsilon_{\beta\delta}
-\epsilon_{\alpha\beta}\epsilon_{\gamma\delta}$
we find that the action naturally splits in two contributions
\ba
& & S \, = \,  S_1 + S_2 \ \ \ \nonumber\\[5.5mm]
\label{tsigm}
S_1 \is  \int \! \rth \, \rtg \, \Bigl(R_h +
 \epsilon_{\alpha\gamma}
\epsilon_{\beta\delta}\nabla^\alpha \Vi^\beta \, \nabla^\gamma V^{i\delta}
 \Bigr)\\[4mm]
S_2 \is
\int \! \rth \, \rtg \, \Bigl(-\hf(\epsilon_\ab \partial^\alpha
\Vi^\beta)^2 \  +  \ b^i \epsilon_\ab \partial^\alpha
\partial_i c^\beta\Bigr) \nonumber.
\ea

\noindent
Notice that, when we insert (\ref{vacuum}) and (\ref{parX}), this action is
highly non-linear in the fundamental fields $X^a$.
It describes, as promised, a topological field theory: there are no
local physical degrees of freedom. The first term is invariant under
local variations of the $X^a$-fields, and can in fact be written as
 a total
derivative. The second term in (\ref{tsigm}), when regarded as a
two-dimensional
 field
theory in the transversal $y$-plane, represents a kind of two-dimensional
 topological
sigma model \cite{wittop}. In this interpretation  the target space of the
model
is given by the space of diffeomorphisms of the $x^\alpha$ plane.

Both terms in the action (\ref{tsigm}) are invariant under the nilpotent BRST
symmetry
\ba
\label{brst}
\delta X^a \is \varepsilon c^\beta \partial_\beta X^a, \nonumber \\[2mm]
\delta b^i \is \varepsilon (\epsilon_\ab \partial^\alpha
V^{i\beta} + c^\beta \partial_\beta b^i), \\[2mm]
\delta c^\alpha \is \varepsilon \, c^\beta \partial_\beta c^\alpha,\nonumber
\ea
under which $\Vi^\alpha$ transforms as
\be
\delta
\Vi^\alpha = \varepsilon (-\partial_i c^\alpha + c^\beta
\partial_\beta\Vi^\alpha-\Vi^\beta\partial_\beta c^\alpha). \ \ \
\ee
This nilpotent symmetry removes all local fluctuations of the $X^a$-field,
and is a remainder of the BRST-invariance of the
original theory, as can be seen as follows.

The original anti-ghost field
$b_{i\alpha}$ transforms under the BRST-symmetry into the
$R_{i\alpha}$ components of the Ricci-tensor, which are the constraints
associated with our gauge choice. Therefore, the
redefinition (\ref{ghredef}) of the ghost field is compatible with
BRST-invariance only if $R_{i\alpha}$ is also of the form
\be
\label{ricciredef}
\ \ \ R_{i\alpha} = \epsilon_\ab \partial^\beta R_i.
\ee
Using that the background satisfies the
field equations of $S_\perp$, which gives that  $\,\nabla^\alpha
R_{i\alpha}=-\nabla^jG_{ij}=0\ $, one deduces that  (\ref{ricciredef}) indeed
holds. Explicitly, one has
\be
\label{ri}
R_i=\epsilon_\ab\partial^\alpha\Vi^\beta.
\ee
The BRST-symmetry of the reduced action (\ref{tsigm}) thus directly follows
from that of the gauge fixed Einstein action. It effectively imposes the
constraint $R_i\! =\! 0$, which tells us that the flows
$\Vi$ are curl free.

Let us end this subsection with one final comment. The attentive reader
may have worried a little about the fact that the changes of
variables from $g_\ab$ to $X^a$ and from $b_{i\alpha}$ to $b^i$, both
require some Jacobians to correctly treat the functional measure.
It turns out, however, that these two Jacobians cancel each other.
Explicitly, if we define the measures of $g_\ab$ and $X^a$ through
\ba
||\delta g_\ab||^2 \is \int\! \rth\rtg \, \epsilon^{\alpha\gamma}
\epsilon^{\beta\delta} \delta g_\ab \, \delta g_\cd, \nonu
||\delta X^a||^2 \is \int\! \rth \rtg \, \eta_{ab} \delta X^a \delta X^b,
\ea
one obtains from (\ref{param})
that the measures of $g_\ab$ and $X^a$ are related via
\be
[dg_\ab] = [dX^a] \, |\det (\epsilon_\ab\nabla^\beta)|^2.
\ee
Similarly, since the ghosts are anti-commuting, we deduce from
(\ref{ghredef}) that
\be
[db_{i\alpha}] = [db_i]\, |\det(\epsilon_\ab\nabla^\beta)|^{-2}.
\ee
So the total Jacobian of the substitution $(g_\ab,b_{i\alpha}) \rightarrow
(X^a,b^i)$ is indeed equal to 1.

\newsection{The Action for the Boundary Values}

Let us now ask ourselves what physical degrees of freedom are described by the
action (\ref{tsigm}). For definiteness, we will from now on consider
the situation where the $x^\alpha$-coordinates parametrize (a part of) the
two-dimensional Minkowski plane $M$. Thus our space has a boundary $\partial M$
in this direction, consisting of four asymptotic regions, two belonging to the
past null infinity ${\cal I}^-$ and two constituting the future null infinity
${\cal I}^+$ (see  fig.\ 1).
Now we notice that local variations of the fields
$X^a$ are not physical, since they correspond to diffeomorphisms and
are cancelled by the BRST-symmetry.
Therefore the only possible physical degrees of freedom are the boundary values
\be
\label{bcond}
 \Xbar^a \equiv X^a{}_{\strut|\partial M}.
\ee
We will allow these boundary values $\Xbar^a$ to be arbitrary, because,
as we will discuss below, this enables us to incorporate
matter into the theory. Furthermore, to make them into true physical
 observables,
we impose the usual restriction that gauge transformations of the theory,
\ie diffeomorphisms, asymptotically go to the identity transformation.
After gauge fixing this translates into the asymptotic condition that the
$c$-ghost vanishes on $\partial M$, and thus the $\Xbar^a$ are BRST-invariant.
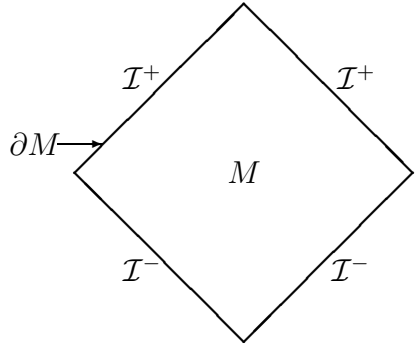
\asympt{Fig.\ 1: The boundary $\partial M$ of the two-dimensional
parameter space $M$ of the
longitudinal coordinates $x^\alpha$ has four components corresponding to
the four asymptotic regions of the Minkowski plane.}

We will now show that for physical field configurations the action
(\ref{tsigm})
can be expressed purely in terms of $\Xbar^a$.
This is most evident for the first term $S_1[X]$: it is easy to see
that it is the integral of a total derivative. Hence
\be
S_1[X]=S_{{}_{\! \partial M}}[\Xbar].
\ee
One finds
\be
\label{faction}
S_{{}_{\partial M}}[\Xbar] =  \oint_{\strut{\partial M}} \!\!\!\!\! dx^\alpha
\int \! \sqrt{h} \, \epsilon_{ab}\Bigl( R_h \Xbar^a \partial_\alpha \Xbar^b
+  \partial_i \Xbar^a \partial_\alpha\partial^i \Xbar^b \Bigr).
\ee
The second term of the action (\ref{tsigm}) plays the role of the gauge fixing
 term to the
action, needed to eliminate the redundancy caused by the topological nature
of the first term. Following a standard argument, often applied in
topological field theory \cite{wittop}, we can use the symmetry (\ref{brst})
of the action to restrict the functional integral to the BRST-invariant field
configurations. This implies that we may put the ghost fields $c^\alpha$
equal to zero, and in addition impose the constraint
\be
\label{eom}
\epsilon_\ab \partial^\alpha \Vi^\beta =\epsilon^\ab\partial_\alpha
(\partial_iX^a\partial_\beta X_a) = 0.
\ee
When we insert this into (\ref{tsigm}) we find that the $S_2$-piece of
the action vanishes identically. We conclude, therefore, that for physical
field configurations the total action indeed reduces to the simple boundary
term
(\ref{faction}).

Finally, a more precise specification of what we mean by the asymptotic
values $\Xbar^a$ of $X^a$
is perhaps in order here, since a priori at least one of them becomes infinite
when we approach the null infinities ${\cal I}^\pm$. The point here
is that, as we will see, only {\it variations} of the asymptotic values of the
$X^a$ are physically relevant.
Hence it is sufficient for our purposes to consider, instead of the
full Minkowski plane, only a finite region $M$, large enough so that variations
$\delta X^a$ at the boundary $\partial M$ are equal to the asymptotic
variations. This procedure will be implicitly adopted in the following
sections.

\newsubsection{Coupling to Matter}

We will now  discuss the coupling of external matter to the
gravitational field, and show that it can also be represented
entirely in terms  of the boundary values $\Xbar^a$. We consider the situation
in which asymptotically all particle momenta
are in the $x^\alpha$-direction, so that  at the boundary $\partial M$
all components except $T_\ab$ vanish.

First let us assume that also in the interior region
$T_\ab$ are the only non-vanishing components.
Its conservation law then reads
\be
\label{conserved0}
\nabla^\alpha T_\ab = 0.
\ee
This tells us that the matter source term $S_m[g]$ is invariant under
diffeomorphisms of $x^\alpha$. In terms of the $X^a$ field variables this
implies that $S_m[g]$ is invariant under local variations, and indeed
only depends on the boundary values $\Xbar^a$. To write this dependence,
it is convenient to represent the incoming and outgoing matter at the
boundary $\partial M$ by means of a momentum flux $P_{a\alpha}$, defined in
terms of the stress-energy tensor via
\be
\label{tp}
T_{\ab} = P_{a\alpha} \partial_\beta X^a.
\ee
Then the variation of the matter contribution with respect
to $X^a$ is
\ba
\label{couple}
\delta S_m[g(X)] \is 2 \int \! \rth\rtg \, \nabla^\alpha (P_{a\alpha}
 \delta X^a)
\nonumber \\[3.5mm]
\is 2 \oint_{\partial M}\!\!\!\!\!  dx^\alpha
\int \!\! \sqrt{h} \, \epsilon_\alpha{}^\beta \, P_{a\beta}\,  \delta \Xbar^a.
\ea
Thus, the matter momentum in- and out-flux couples directly to the variations
of the boundary values $\Xbar^a$.

In general, however, it is possible that in the interior of $M$
also the mixed components $T_{i\alpha}$ of the stress tensor
become non-vanishing. Although in this case we have to refine the above
derivation of the matter coupling, it in fact turns out that the final
result (\ref{couple}) remains unaltered. Let us indicate how this
goes. The conservation equation (\ref{conserved0}) is now modified to
\be
\label{conserved}
\nabla^\alpha T_\ab + \nabla^i T_{i\beta} = 0 .
\ee
Because we assume $T_{ij} = 0$, we know that $T_{i\alpha}$ in
addition satisfies $\nabla^\alpha T_{i\alpha} = 0$. It is therefore
of the form
\be
\label{tj}
T_{i\alpha} = \epsilon_\ab \partial^\beta J_i.
\ee
Here $J_i$ can be identified with the current associated with the boosts
$\delta X^a=\epsilon^{a}_bX^b$ in the $X$-plane. Namely,
by integrating (\ref{conserved}) over the plane $M$ against
$\xi^\beta\!\! =\! \partial^\beta X^a X^b \epsilon_{ab}$ and inserting the
redefinitions
(\ref{tp}) and (\ref{tj}) we deduce that for all $y$
\be
\label{jcons}
\nabla_i \int_M \!\! \rtg \,  J^i \, =
\oint_{\partial M} \!\!\!\! dx^\alpha  \epsilon_\alpha{}^\beta P_\beta^{a}\,
\Xbar^b\epsilon_{ab},
\ee
which expresses the conservation of angular momentum.

When $J_i$ is non-vanishing, the matter action
$S_m[g]$ is no longer invariant under local variations of $X^a$.
However, coordinate invariance tells us that
this local $X^a$-dependence must cancel when we include the gravitational
interaction. To see this, consider the BRST-variation of the matter
action $S_m[g]$. It can be expressed in terms of the current $J_i$ as
follows
\ba
\delta_{brst} S_m[g] \is 2\varepsilon\int \! \rth\rtg \, T_\ab \nabla^\alpha
 c^\beta
 \nonumber
\\[3.5mm]
\is 2 \varepsilon \int \! \rth \rtg \,  J^i \, \epsilon_\ab \partial^\alpha
 \partial_i
c^\beta,
\ea
where we used (\ref{conserved}) and (\ref{tj}). Thus the combined matter and
gravitational action is BRST-invariant, provided we include $J^i$ in the
transformation law (\ref{brst}) of the anti-ghost. As a consequence the
constraint (\ref{eom}) on physical field configurations is replaced by
\be
\label{constraint}
\epsilon_\ab \partial^\alpha \Vi^\beta = J_i.
\ee
With this modification we can now follow the argument of the preceding
 subsection
to show that, under the above physical constraint, the dependence of $S_m[g]$
on
local variations of $X^a$ indeed cancels against that of the second term of the
gravitational action (\ref{tsigm}). Hence also in this case the entire coupling
to matter is described by equation (\ref{couple}).

\newsection{The Scattering Matrix}

In this section we will determine the amplitudes describing the gravitational
scattering of the matter particles. First we like to
summarize the result of the above analysis. We have shown that, due to
the BRST-invariance,
the functional integral over the fields $X^a$, $b^i$, and $c^\alpha$ is
 concentrated
on the space of physical field configurations, satisfying the constraint
(\ref{constraint}). For given boundary values $\Xbar^a$, this constraint
uniquely fixes the interior values of $X^a$, and thus all dynamics
is reduced to the boundary $\partial M$.
It will be convenient to introduce a `time'-variable $\tau$ which
parametrizes the coordinates $x^\alpha(\tau)$ on $\partial M$.
Notice that, since $\partial M$ is closed, $\tau$ is periodic.
The action that describes the dynamics of the boundary values and its coupling
 to matter is
\be
\label{baction}
S_{{}_{\!\partial M}}[X]= S[X] +
2 \int \! d\tau\! \int\!\! \sqrt{h}\,\,P_{a,\tau} X^a,
\ee
with
\be
\label{raction}
S[X] =
 \int \! d\tau\int\!\sqrt{h}\,\,
\epsilon_{ab} {\partial X \over\partial\tau}^a (\Delta_h-R_h)X^b,
\ee
where $\Delta_h$ denotes the scalar Laplacian in the transversal $y$-plane.
Here we used (\ref{couple}) to represent the matter coupling in terms of the
momentum density
$
P_{a,\tau} = \epsilon^\alpha_\beta  P_{a\alpha} \partial_\tau x^{\beta}.
$
In (\ref{baction}) and from now on we drop the bar on the boundary values of
 $X^a$. In the rest of this section it is understood that
all the fields are defined on $\partial M$.

\newsubsection{Quantum mechanics of the boundary values.}

The boundary theory (\ref{baction}) can naturally be
regarded as a (non-relativistic) 2+1-dimensional field theory.
Indeed to obtain the scattering amplitudes we will still need to
perform the remaining functional integral over the $X^a$, and
it is therefore a natural step to consider the quantum mechanics of
this boundary theory. For this it is a fortunate fact that
the action (\ref{raction}) is quadratic in the fields $X^a$, especially
since it was derived from the highly non-linear Einstein action.
Canonical quantization is therefore straightforward. It
leads to the following commutation relations
\be
\label{commutator}
[X^a(y_1),X^b(y_2)] = i \epsilon^{ab} f(y_1,y_2),
\ee
where
\be
\label{green}
(\Delta_h - R_h)f(y_1,y_2) = \delta^{(2)}(y_1,y_2)
\ee
defines the Green function of the modified Laplacian
$(\Delta_h-R_h)$. This commutator
previously appeared in the work of 't Hooft \cite{thooft2}.
Intuitively it can be understood from the Einstein equation (\ref{eineqn}),
which relates the coordinate fields $X^a$ to their canonical
conjugate momenta. Let us denote by ${\cal H}$ the Hilbert space
in which the commutator algebra (\ref{commutator}) is realized;
vectors in ${\cal H}$ can be represented as functionals of $X^+(y)$ or
of $X^-(y)$. Note that, since the action is invariant under
reparametrizations in $\tau$, there is no Hamiltonian, i.e. $H=0$.

The matter coupling is described in terms of `vertex operators' of the form
\be
\label{pexp}
V(P)={\rm{Pexp}}\Bigl( 2i\! \int\!
d\tau\!\! \int\!\! \sqrt{h}\,P_{a,\tau}(\tau,y)X^a(y)\Bigr).
\ee
Here the path-ordering is necessary, as the $X^a(y)$ do not mutually
commute. The momentum flux  $P_{a,\tau}$ will
for the moment still be treated as a classical quantity, but in the
next subsection, when we study the scattering matrix, $P_{a,\tau}$ becomes a
quantum mechanical operator in the Hilbert space of the matter theory.
The idea is that in this way, by first computing the expectation value
of the general vertex operator (\ref{pexp}),
we obtain a generating functional for all scattering amplitudes.
Since the boundary time $\tau$ is periodic, this expectation value
is given by the trace over the Hilbert space ${\cal H}$
\be
\Bigl\langle V(P) \Bigr\rangle
= \tr_{\strut\cal H} \Bigl({\rm{Pexp}}\Bigl[2i\!\oint\!
d\tau\!\!\int\!\!\sqrt{h}\,P_{a,\tau}X^a\Bigr]\Bigr).
\ee
The result can be computed from the commutation relations
(\ref{commutator}) via a simple application of the CBH-formula, or
alternatively, by using the relation with the functional integral.
One finds
\be
\label{gensca}
\Bigl\langle V(P) \Bigr\rangle =
 \exp \Bigl[
i\! \int_{\tau<\tau^\prime}\!\!\!\!\!\!\!\! d\tau d\tau^\prime \int\!
\sqrt{h}\,\epsilon^{ab}P_{a,\tau}(\Delta_h-R_h)^{-1}P_{b,\tau^\prime}
\Bigr].
\ee
In addition we obtain a delta-function imposing the condition
that the total momentum flux at the transversal position $y$ vanishes
\be
\oint d\tau P_{a,\tau}(\tau,y) = 0.
\ee
This condition, which is due to the invariance of $S[X]$ under
$\tau$-independent shifts of $X^a$, guarantees that the result (\ref{gensca})
does not depend on the base-point of the $\tau$-integration.

The result (\ref{gensca}) for the generating function is exactly
given by the semi-classical expression, $\exp(iS[X_{cl}])$, where
$X_{cl}$ solves the classical equation of motion
\be
\label{eineqn}
(\Delta_h-R_h){\partial X^a\over \partial\tau}=
\epsilon^{ab} P_{b,\tau}.
\ee
This equation, which is in fact the remainder of the Einstein equation,
only involves the values of the coordinate fields $X^a$ and momenta $P_a$ on
the
boundary $\partial M$. Thus with our formalism we have succeeded in computing
the Einstein action of the relevant classical solution, without needing to
integrate the equation of motion for the interior region.

In principle the classical configuration for $X^a$ in the interior
region is determined by the physical constraint (\ref{constraint}) involving
the current $J_i$.  The integral
of (\ref{constraint}) over the two-dimensional parameter space $M$
gives for all $y$ the relation
\be
\label{integra}
\oint_{\partial M} \!\!\!\! d\tau \, {\partial
X\over\partial\tau}^a\partial_iX_a(y) = \int_M \!\! \rtg \, J_i(y).
\ee
where we used that
 $\epsilon_\ab\partial^\alpha\Vi^\beta=\epsilon^\ab\partial_\alpha
(\partial_iX^a\partial_\beta X_a)$.
As a consistency check one can verify, using the equation of motion
 (\ref{eineqn}),
that this relation implies the conservation law (\ref{jcons}).

\newsubsection{Description of the Scattering Matrix}

Let us now determine the scattering matrix for (point-like) matter particles.
Since we assume their masses to be negligible compared to the Planck mass, they
are divided into left-movers with momenta $p_-^i$ and right-movers with momenta
$p_+^j$. Now, it is important to note that the left- and right-movers come in
from or go out to different asymptotic regions of space-time,
so that the boundary $\partial M$ is divided into four different components,
on each of which either $P_{-,\tau}$ or $P_{+,\tau}$ vanishes, (see fig.\ 2).
Correspondingly, the vertex operator (\ref{pexp}) representing the momentum
flux
decomposes into the product of four vertex operators, where, for example, the
operator representing the right-moving incoming particles has the form
\figuur{0}{Fig.\ 2: The amplitude (\ref{vvvv}) is the $S$-matrix element
between the
state with momentum flux $P_a^{in}$ at the in-region ${\cal I}^-$ and the state
of given out-going flux $P^{out}_a$ at ${\cal I}^+$.}
\be
V(\Pip)=\exp\Bigl( 2i\! \int \! \sqrt{h}\,
\Pip(y)X^+(y)\Bigr),
\ee
where
\be
\Pip(y) =  \int_{{\cal I}^-}\!\!\! d\tau P_{+,\tau}(\tau,y)
\ee
is the total incoming momentum at the transversal position $y$.
The scattering amplitude is obtained by specializing the general formula
(\ref{gensca}) to the present situation. This gives the result
\be
\label{vvvv}
\Bigl\langle V(\Pip)\, V(\Pim) \, V^*(\Pop) V^*(\Pom)\Bigr\rangle
= S(P) \, \delta\Bigl(P_a^{out} - P_a^{in} \Bigr),
\ee
where
\be
\label{scat}
S(P) = \exp\Bigl(i \int\!\!  \sqrt{h} \! \int\!\!   \sqrt{h}
\, \eta^{ab} P_a(y_1) f(y_1,y_2) P_b(y_2)\Bigr)
\ee
is the scattering matrix.

The variables $P_a(y)$ act as quantum mechanical operators in the matter
Hilbert space, and can be used to characterize the in- and out states.
Because the operators $P_a(y)$ mutually commute, we can find a basis of
eigenstates in the in- and out-Hilbert space. Such a basis is obtained by
specifying for each particle its longitudinal momentum $p_a^{(i)}$ and
transversal coordinate $y^{(i)}$. On the $N$-particle state
$|p^{(i)},y^{(i)}\rangle$,  the momentum operator $P_a(y)$ has
the eigenvalue
\be
P_a(y)\,|p_a^{(i)},y^{(i)}\rangle
= \sum_{i=1}^{N} p_a^{(i)} \delta(y,y^{(i)})\,|p^{(i)},y^{(i)}\rangle.
\ee
Since the $S$-matrix (\ref{scat}) is expressed purely in terms of
the $P_a(y)$, it is also diagonal in the same basis, with eigenvalues
\be
\label{scamp}
S(p^{(i)},y^{(i)})
= \prod_{i\neq j} \exp\Bigl( i \eta^{ab} p_a^{(i)}p_b^{(j)}
f(y^{(i)},y^{(j)}) \, \Bigr).
\ee
Equations (\ref{scat}) and (\ref{scamp}) give the final result for
the scattering matrix. Notice that the $N$-particle amplitude
factorizes into the product of two-particle amplitudes, which tells
us that there are no three- or higher point interactions. In the next section
we will discuss the two-particle process in more detail.

We observe that the $S$-matrix element (\ref{scamp}) can be
written in a form similar to a string amplitude
as follows
\be
S(p^{(i)},y^{(i)}) =
\int\! [dX]
\,e^{iS_{str}[X]} \, \prod_i e^{2ip_a^{(i)} X^a(y^{(i)})},
\ee
where $S_{str}$ denotes the two-dimensional action for a string embedded in
a $1+1$ dimensional space-time (cf.\  \cite{thooft2})
\be
S_{str}[X] =\int\! d^2 y \sqrt{h}\Bigl(h^{ij}
\partial_iX^a\partial_jX_a+R_hX^a X_a\Bigr).
\ee
An unusual feature is the factor $i$ in front of the action,
even though the `world sheet', the $y$-plane, has Euclidean signature.

Another difference with the usual string amplitudes is that there is
no integration
over the positions of the vertex operators, nor over the metric $h_{ij}$.
The reason for not integrating over these variables is that we assume
that for large $x^\alpha$ we are in the classical regime, where
general covariance is broken in the $y$-direction
via some vacuum expectation value
$\langle h_{ij}\rangle = h_{ij}$. One can imagine, however, some physical
situations where general covariance remains unbroken. In this case
we {\it do} have to integrate over $h_{ij}$  and the formal analogy with
string theory becomes almost exact.

\newsection{The Two-Particle Amplitude}

In this section we will take a more detailed look at the two-particle
scattering process, and discuss its geometrical interpretation.
In the following we will assume that the transversal metric is
flat $h_{ij}=\delta_{ij}$. In this case the Green function $f(y_1,y_2)$
becomes
\be
f(y_1,y_2) = -\log |y_1-y_2|^2.
\ee

\newsubsection{The Shock Wave Geometry}

The scattering matrix obtained in the previous subsection
has a concrete physical interpretation
in terms of gravitational shock waves \cite{thooft1,shock}.
To exhibit this interpretation, we first observe that the N-particle
amplitude factorizes into the product of two-particle amplitudes,
given by
\be
\label{s12}
S_{12} = \exp\Bigr( 2 i p^-_1 p^+_2 f(y_1,y_2) \Bigr)
\ee
where $p^\pm_i$ and $y_i$ are the momentum and transversal coordinate of
particle $i$. This two-particle result is obtained from (\ref{scat})
by choosing the in-flux of momentum at ${\cal I^-}$ of the following form
\ba
\label{pin}
P_{in}^-(x,y) \is  p^-_1 \, \delta(x^+\!\!\! -\! x_1^+) \, \delta(y \! -\!
y_1),
 \nonu
P_{in}^+(x,y) \is   p^+_2 \, \delta(x^-\!\!\! -\! x_2^-) \,
\delta(y \! - \! y_2).
\ea
The corresponding classical solution of (\ref{eineqn}) is given by
\ba
\label{class}
X_{in}^-
\is x^-
- \, p^-_1 \, \theta(x^+\!\!\!-\!x^+_1)\, f(y,y_1),\nonu
X_{in}^+
\is x^+
- \, p^+_2 \, \theta(x^-\!\!\!-\!x^-_2)\, f(y,y_2).
\ea
These classical field
configurations for $X_{in}^+$ and $X_{in}^-$ both exhibit a discontinuity
at the $x^\alpha$ trajectory of one of the particles. These discontinuities
are the well-known gravitational impulsive wave, or `shock wave',
solutions to the Einstein equation obtained in \cite{shock}.
The form of this shock wave geometry for a single
particle is indicated in fig.\ 3. It shows that a right-moving geodesic is
 shifted
by an amount $p^-_1 f(y,y_1)$ when it passes through the $x^\alpha$-trajectory
of the fast left-moving particle with momentum $p_1$.
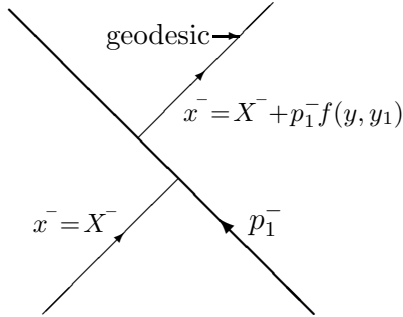
\oneshock{Fig.\ 3: The shock wave geometry of a fast left-moving particle
with momentum $p_1^-$ and transversal position $y_1$ has for all $y$ a
discontinuity at the $x^\alpha$-trajectory of the particle.}

The formula (\ref{class}) only describes the two gravitational shock waves
in the in-region, \ie before they collided. It is clear that
in the out-region the solution will become more complicated. To solve
for the complete metric describing the two colliding shock waves indeed
appears to be a rather non-trivial problem. The technical difficulty here
is that after the two particle trajectories cross each other, the
 position of each shock wave is shifted by the discontinuity of the other.
This distorts the waves in such a way that they are
no longer stable solutions of the Einstein equation.
What we know, however, is that the
out-flux $P_{\tau,a}(\tau,y)$ at ${\cal I^+}$ must, by momentum conservation,
 yield the same integrated momentum distribution $P_a(y)$ as the
in-flux. This is all information we need to compute
the scattering amplitude $S_{12}$ via the general formula (\ref{scat}).

Finally, let us note that the amplitude (\ref{s12}) has been derived
in an alternative way in \cite{thooft1}
by considering the effect of the shock wave
of one of the two colliding particles on the wave packet of the
other.  This reasoning makes use of the
fact that one can go to a frame where one of the particles
moves slowly, so that only the gravitational field of the other
fast moving particle needs to be considered. As an operator in
the Hilbert space of the slow particle, the $S$-matrix (\ref{s12})
can then be recognized as the generator
of the coordinate shift caused by the shock wave of the fast particle.

\newsubsection{Partial Wave Analysis}

The two-particle process is special because in this case our geometrical setup
in fact does not break Lorentz invariance: the division into longitudinal and
transversal directions puts no restrictions on the momenta of the two
particles,
but is defined in terms of them. We will now indeed show that the
two-particle $S$-matrix and differential cross section can be
given in a Lorentz invariant form.
Our approach is based on a partial wave analysis of the scattering amplitude,
and can be regarded as a refinement of the method of \cite{thooft1}. It was
found in \cite{thooft1}  that the two-particle amplitude possesses an
infinite number of poles for imaginary $P^2$, which were conjectured to
indicate the presence of bound states. One of our aims is to determine
whether these poles already occur at the level of partial waves or not.

To write the $S$-matrix
(\ref{s12}) in manifestly Lorentz invariant form, we can first
replace $p_1^+p_2^-$ by $p_1^\mu p_{2\mu}$. Next, let us
introduce the total linear and angular momentum operators
\ba
P^\mu \is p_1^\mu + p_2^\mu, \\[3mm]
J_{\mu} \is {1\over |P|} \epsilon_{\mu\nu\lambda\sigma}
 p^\nu_1 p_2^\lambda(x_1^\sigma-x^\sigma_2),
\nonumber
\ea
and express $p_1^\mu p_{2\mu}$ and the impact parameter $|y_1-y_2|$ in
 terms of these quantities as (here we use that $p_1^2 = p_2^2 = 0$)
\ba
2p_1^\mu p_{2\mu} \is P^2, \\[2.5mm]
|y_1-y_2|^2 \is J^2/P^2.
\ea
Notice that this implies that as a quantum mechanical operator
the impact parameter has a discrete spectrum.

The two-particle scattering operator is conveniently expressed in terms of the
mutually commuting operators $P$ and $J$ as follows
\be
S_{12} = \left( J^2\over P^2\right)^{-i P^2}.
\ee
Hence the S-matrix is diagonal in the partial waves basis of eigenmodes
$|P,l,m\rangle$ of $P$, $J^2$ and $J_3$, and can be represented as
\be
S_{12} | P,l,m\rangle =
e^{2i\delta_{l}(P)} |P,l,m\rangle,
\ee
where the phase shift $\delta_l(P)$ is given by
\be
\label{phase}
\delta_l(P) = -\hf P^2 \log\Bigl({l(l+1)\over P^2}\Bigr).
\ee
This result gives the exact leading behaviour of the scattering matrix
for large angular momentum $l$, that is
\be
l(l+1) \gg P^2,
\ee
with $P^2$ measured in Planck units.
Notice therefore that if $P^2$ is small our approximation is
reliable for all partial waves, except for the
$s$-wave. Indeed, the above result for the phase shift becomes singular
for $l=0$. Notice further that, for fixed $l$, the $S$-matrix does not
exhibit any poles in $P^2$.

{}~From the knowledge of the phase shifts $\delta_l(P)$ one can compute the
differential cross-section via
\ba
{d\sigma\over d\theta}(\theta) \is |f_P(\theta)|^2 \nonu
f_P(\theta) \is {1\over2 i |P|}
\sum_{l=0}^\infty (2l+1) \, (e^{2i\delta_l(P)}\! - \! 1 )
 P_l(\cos\theta).
\label{fpsum}
\ea
Inserting the result (\ref{phase}) (and restricting the sum in (\ref{fpsum}) to
$l\!>\!0$) we obtain an expression for the amplitude $f_P(\theta)$ which is
reliable for small scattering angles $\theta$. To extract this leading
behaviour
we make use of the following asymptotic formula, which becomes exact
for large $l$ and small $\theta$
\be
P_l(\cos \theta)\rightarrow J_0\Bigl((2l\!+\!1)\sin{\hf \theta}\Bigr),
\ee
where $J_0$ is the first Bessel function. Let us now
re-introduce the impact parameter $y=\sqrt{l(l\!+\!1)}/|P|$ and
approximate the sum over $l$ by the integral over $y$.
The resulting integral can be performed and gives in this
approximation the following leading behaviour
\ba
\label{fp}
\qquad \qquad \qquad f_P(\theta)\! & \! =\! &\! -i |P|\int_0^\infty \!\! dy\,
 y^{1-2iP^2}
J_0\Bigl(2 y\,|P| \sin{\hf\theta}\Bigr) \quad \qquad \qquad {\mbox{($\theta$
 small)}} \nonu
\! & \! =\! &\! {|P|\over 2\sin^2{\hf \theta}}\,
{\Gamma(1- {i} P^2) \over \Gamma(1+{i}P^2) }\,
\Bigl(P^2 \sin^2 {\hf\theta}\Bigr)^{{i}P^2}.
\ea
This formula agrees with the result obtained in \cite{thooft1}.

The most striking feature of the scattering amplitude (\ref{fp}) is the
occurrence of poles at the center of mass energies $P^2=-iN$, in Planck units.
However, it is clear from the above derivation that they do not correspond to
any resonant states of quantum gravity, but are produced by
the replacement of the sum over $l$ by the integral over $y$.
If there would exist resonances with definite spin,
the corresponding poles would already have been visible in the phase
shifts $\delta_l(P)$.
The poles in (\ref{fp}) come from the $y\rightarrow 0$ part of the integral
and disappear when this part of the integral is cut-off and replaced by a sum.
 Hence, assuming that the $s$-wave contribution
has no poles, the original expression (\ref{fpsum}) for the
scattering amplitude is analytic for all values of $P^2$.
We conclude therefore that result (\ref{fp}) describes the correct small
$\theta$ behaviour of the amplitude as long as Im$P^2>-1$; beyond this point
subleading terms in (\ref{fpsum}) become important.

\newsection{Concluding Remarks}

In this paper we have described a systematic method for studying
forward scattering in quantum gravity. It exploits the presence
of the small dimensionless parameter $g_\perp=(l_{pl}/l_\pper)^2$, in
terms of which one can do perturbation theory. The final result (\ref{scat})
for the scattering amplitude is valid only to leading order in $g_\perp$,
but is {\it non-perturbative} in the other dimensionless coupling
constant $g_\ppar=(l_{pl}/l_\ppar)^2$. It can therefore be applied for
arbitrarily large center of mass energies.

We have shown that the relevant gravitational modes that mediate between
the high energy particles are described by means of a topological field
theory and that their dynamics and the coupling to matter is
formulated in terms of the boundary values of $X^a$.
This result is universal in the sense that it only depends on
general covariance and dimensional analysis, and also applies in the
presence of a cosmological constant or curvature squared terms.
Curvature squared terms will, however, modify the boundary action
 (\ref{faction})
by higher derivative terms. The fundamental property that remains is the
absence of local dynamics at short distances. This is a reflection of the fact
that in the $x^{\alpha}$-direction general covariance in unbroken, and in
this sense we have truly dealt with the `topological phase' of
quantum gravity.

The fact that the leading order amplitude (\ref{scat}) is unitary, tells
us that to this order no gravitational radiation is produced in the
scattering process. Indeed, the coupling to gravitons is of lower order
in the momentum transfer. It is possible to compute these higher
order corrections due to the presence of gravitons,
although at some point one is bound to encounter the problem
of non-renormalizability of quantum gravity. String models of
quantum gravity are likely to avoid this problem, but we expect
that these string modifications will not drastically alter the
leading order result (see, however, \cite{amati}).

One of the more intriguing conclusions of our paper is the commutation
relation (\ref{commutator}) between the coordinate fields $X^a$.
Its geometrical origin lies in the fact that the discontinuities describing
the gravitational shock waves (\ref{class}) of the left- and right-moving
particles are in a way incompatible: the shift caused by the left-mover
causes an uncertainty in the location of the right-mover, and vice versa,
at the point where the two $x^\alpha$ trajectories intersect.
One may speculate that the commutator hints towards the existence of a new
quantum gravitational uncertainty principle, which could have radical
consequences. Indeed, 't Hooft has proposed that, by taking this uncertainty
principle serious, one can construct an $S$-matrix for the successive
black hole formation and evaporation process \cite{thooft2}.

In view of this application, it would be very interesting to generalize
our analysis to the case where $x$, instead of being a cartesian coordinate,
represents a radial coordinate $r$. This corresponds to the physical situation
in which the momenta of all particles are directed to the same point
($r\!=\!0$) in space. However, here we will meet a new difficulty, since for
very small values of $r$ our assumption about the transversal momenta being
small becomes unattainable. Therefore additional modes of the gravitational
field need to be included in the analysis. One indeed expects that in
this situation black hole formation may take place, and then \eg the
mass of the black hole becomes a dynamical variable.

\renewcommand{\thesection}{A}
\renewcommand{\thesubsection}{A.\arabic{subsection}}

\vspace{15mm}
\pagebreak[3]
\setcounter{section}{1}
\setcounter{equation}{0}
\setcounter{subsection}{0}
\setcounter{footnote}{0}
\begin{center}
{\sc Appendix: The Einstein Action in the Gauge (2.2)}
\end{center}

\medskip

In this Appendix we outline the derivation of the formula
(\ref{s1})-(\ref{s2}) for the Einstein action in the
gauge
\be
\label{a1}
g_{\mu\nu} = \left( \begin{array}{cc} g_\ab & 0 \\ 0 & h_{ij}
\end{array} \right)
\ee
Here and in the following $\mu, \nu$ run from $0$ to 3, $\alpha,\beta$ from
$0$ to 1  and $i,j$ from 2 to 3.
Under this decomposition the Ricci scalar of $g_{\mu\nu}$ reads
\be
\label{a2}
R^{(4)} = g^\ab g^\cd R_{\alpha\gamma\beta\delta} + h^{ij} h^{kl}
R_{ikjl} + 2 g^\ab h^{ij} R_{\alpha i \beta j}
\ee
where $R_{\mu\nu\lambda\sigma}$ is the Riemann tensor
\be
\label{riem}
R_{\mu\nu\lambda\rho} = \partial_\lambda \Gamma_{\mu\rho\nu}
- \partial_\rho \Gamma_{\mu\lambda\nu} + \Gamma_{\mu\lambda\sigma}
\Gamma^\sigma_{\rho\nu}-
\Gamma_{\mu\rho\sigma}\Gamma^\sigma_{\lambda\nu}
\ee
Inserting the explicit form of the following Christoffel symbols
\ba
\Gamma_{i\ab} =  &\! -\Gamma_{\ab i} \! & \!  = - \hf \partial_i g_\ab
\nonu
\Gamma_{\alpha ij} =\!  &\! -\Gamma_{ij\alpha} \! & \!  =
- \hf \partial_\alpha h_{ij}
\ea
while keeping $\Gamma^\gamma_\ab$ and $\Gamma^k_{ij}$ implicit, we find for the
first two terms in (\ref{a2})
\ba
g^\ab g^\cd R_{\alpha\gamma\beta\delta} \is R_g - \, {\textstyle{1\over 4}}
h^{ij} \, \partial_i g_\ab
\, \partial_j g_\cd \, (g^{\alpha\delta}g^{\beta\gamma}\! -\! g^\ab g^\cd)
 \nonumber\\[3.5mm]
h^{ij} h^{kl} R_{ikjl} \is R_h
- \, {\textstyle{1\over 4}}\,  g^\ab \,
\partial_\alpha h_{ij} \, \partial_\beta h_{kl}
\, (h^{il} h^{jk}\! -\! h^{ij}h^{kl})
\ea
where $R_g$ and $R_h$ are the scalar curvatures of $g$ and $h$ considered as
two-dimensional metrics. The last term in (\ref{a2}) becomes
\ba
2 g^\ab h^{ij} R_{\alpha i \beta j} \is
\ \ \hf\,  h^{ij} \, \partial_i g_\ab \,
\partial_j g_\cd \, (g^{\alpha\delta}g^{\beta\gamma}\! -\! g^\ab g^\cd)
 \nonumber\\[3.5mm]
& & + \, \hf \, g^\ab \, \partial_\alpha h_{ij} \, \partial_\beta h_{kl}
\, (h^{il} h^{jk}\! -\! h^{ij}h^{kl})\\[3.5mm]
& & + {1\over \sqrt{g}}\nabla^i\partial_i \sqrt{g} + {1\over \sqrt{h}}
\nabla^\alpha \partial_\alpha \sqrt{h}\nonumber
\ea
Here the last two terms contribute total derivatives to the Einstein action,
and
can therefore be omitted. Adding the remaining terms and using
\ba
g^{\alpha\delta} g^{\beta\gamma} \! -\! g^\ab g^\cd \is
\epsilon^{\alpha\gamma}\epsilon^{\beta\delta} \nonumber\\[3mm]
h^{il} h^{jk}\! -\! h^{ij}h^{kl} \is \epsilon^{ik}\epsilon^{jl}
\ea
leads to the result (\ref{s1})-(\ref{s2}) for the Einstein action.

{\renewcommand{\Large}{\normalsize}
}

\end{document}